\documentclass[preprint]{aastex631}

\usepackage{color, colortbl}
\definecolor{Gray}{gray}{0.9}

\received{June 30, 2021}
\accepted{August 18, 2021}

\submitjournal{PSJ}

\shorttitle{Taurid stream \#628: a reservoir of large cometary impactors}
\shortauthors{Devillepoix et al.}
\graphicspath{{./}{figures/}}

\begin{document}

\title{Taurid stream \#628: a reservoir of large cometary impactors}

\correspondingauthor{Hadrien A. R. Devillepoix}
\email{hadrien.devillepoix@curtin.edu.au}

\newcommand{\curtin}{School of Earth and Planetary Sciences, Curtin University, Perth WA 6845, Australia}
\newcommand{\seti}{SETI Institute, 189 Bernardo Avenue, Mountain View, CA 94043, USA}
\newcommand{\ames}{NASA Ames Research Center, Mail Stop 241-11, Moffett Field, CA 94035, USA}

\author[0000-0001-9226-1870]{Hadrien A. R. Devillepoix}
\affiliation{\curtin}

\author[0000-0003-4735-225X]{Peter Jenniskens}
\affiliation{\seti}
\affiliation{\ames}

\author[0000-0002-4681-7898]{Philip A. Bland}
\affiliation{\curtin}

\author[0000-0003-2702-673X]{Eleanor K. Sansom}
\affiliation{\curtin}

\author[0000-0002-8240-4150]{Martin C. Towner}
\affiliation{\curtin}

\author[0000-0003-4766-2098]{Patrick Shober}
\affiliation{\curtin}

\author{Martin Cup\'ak}
\affiliation{\curtin}

\author[0000-0002-5864-105X]{Robert M. Howie}
\affiliation{\curtin}

\author[0000-0002-8646-0635]{Benjamin A. D. Hartig}
\affiliation{\curtin}

\author[0000-0002-8914-3264]{Seamus Anderson}
\affiliation{\curtin}

\author[0000-0002-0363-0927]{Trent Jansen-Sturgeon}
\affiliation{\curtin}

\author{Jim Albers}
\affiliation{\seti}

\begin{abstract}

The Desert Fireball Network observed a significant outburst of fireballs belonging to the Southern Taurid Complex of meteor showers between October 27 and November 17, 2015. At the same time, the Cameras for Allsky Meteor Surveillance project detected a distinct population of smaller meteors belonging to the irregular IAU shower \#628, the s-Taurids. While this returning outburst was predicted and observed in previous work, the reason for this stream is not yet understood. 2015 was the first year that the stream was precisely observed, providing an opportunity to better understand its nature.
We analyse the orbital elements of stream members, and establish a size frequency distribution from millimetre to metre size range.

The stream is highly stratified with a large change of entry speed along Earth's orbit.
We confirm that the meteoroids have orbital periods near the 7:2 mean-motion resonance with Jupiter. The mass distribution of this population is dominated by larger meteoroids, unlike that for the regular Southern Taurid shower. The distribution index is consistent with a gentle collisional fragmentation of weak material.

A population of metre-sized objects is identified from satellite observations at a rate consistent with a continuation of the size-frequency distribution established at centimetre size. The observed change of longitude of perihelion among the s-Taurids points to recent (a few centuries ago) activity from fragmentation involving surviving asteroid 2015TX\d{24}. This supports a model for the Taurid Complex showers that involves an ongoing fragmentation cascade of comet 2P/Encke siblings following a breakup some 20,000 years ago.

\end{abstract}


\section{Introduction} \label{sec:intro}

The Southern and Northern Taurid showers are part of a Taurid Complex of meteor showers with daytime and nighttime components, the night-time showers of which spread from September to December along Earth's path \citep{jenniskens2006meteor,2013M&PS...48..270B}.  \citet{whipple1940photographic} first identified comet 2P/Encke as the likely parent body. The comet now has evolved to a phase in the rotation of the nodal line that keeps its nodes far from Earth. However, this Jupiter family comet moves in a short 3.3-year orbit that is decoupled from Jupiter, which makes both the comet and meteoroid orbits relatively stable for long periods of time. The wide dispersion of the showers' longitude of perihelion requires a formation age at least 20,000 years ago, the minimum time it takes to disperse the longitude of perihelion of the orbits as wide as observed.

\citet{1984MNRAS.211..953C} first suggested that the a large number of potential other parent body asteroids were part of a Taurid Complex that originated from a giant comet breakup 20,000 years ago. However, \citet{jenniskens2006meteor} pointed out that these early proposed parent bodies appeared to be S- or O-class stony asteroids, instead, which evolved into Encke-like orbits from a source in the asteroid belt via the $\nu_6$ resonance. The same conclusion was also reached more recently by \citet{2014A&A...572A.106P} and \citet{2015A&A...584A..97T}.

\citet{jenniskens2006meteor} and \citet{2016Icar..266..384J} also noticed that there was no mirror image between Taurid shower component nodes
in northern and southern branches, suggesting that meteoroids did not survive long enough to fully disperse their nodes around the original orbit.  A full precession of the nodal line takes about 5,000 years, hence individual shower components in the South and North branches are likely younger than 5000 years. Instead, \citet{jenniskens2006meteor} proposed that a more restricive set of possible parent bodies with semi-major axis close to the 2.22 AU of comet 2P/Encke was responsible and the 20,000 year old stream now reflects the current dispersion of these smaller parent bodies that continue to generate Taurid meteoroids in the recent past. One possible parent was identified as asteroid 2004\,TG10, now known to be a 1.3\,km large object (H = 19.4) with low 0.018 albedo.

This idea that the Taurid complex is active as a whole, and is not just the remnant of a single 20,000+ years old break up, is supported by the orbital analysis done by \citet{1952HelOB..41....3W}. Long before modern orbital integrators and the introduction of orbital similarity criteria $D$ \citep{1963SCoA....7..261S,1981Icar...45..545D}, they were able to identify a group of Southern Taurids that dynamically converged 1400 years in the past. To explain why Encke did not match the orbit of the group, they suggested that the stream of material could have come from a companion, which could have itself separated from Encke earlier. More recently, \citet{2016MNRAS.461..674O} reported two large bolides entering the skies of Poland on October 31, 2015. The meteoroids have very similar orbits ($D_D = 0.011$), and the authors identify two asteroids (2005 UR and 2005 TF50) as potential members of the stream. Using a backward integration, they show that these 4 objects (two meteoroids and two NEOs) have their orbital elements converge 1500 years ago, in good agreement with \citet{1952HelOB..41....3W}. That does not exclude that the bolides originated from one of the two asteroids in more recent times.

The 2015 bolides were part of an outburst of fireballs that is a repeating phenomenon for the Southern Taurid complex in late October and early November. Every 3+ years, there is a significant uptick of Taurid fireballs (Table \ref{table:swarmyears}). There is no clear link to the times when comet Encke returns to perihelion. Instead, it appears that a cloud of meteoroids remains concentrated around a certain position (range of mean anomaly) along the orbit. \citet{2017MNRAS.469.2077O} reported on Taurids observed in the Polish Fireball Network in 2005 and 2015 and found that over 100 fireballs moved in similar orbits as asteroid 2015 TX24. Similarly,  \citet{2017A&A...605A..68S} found that 113 out of 144 Taurid fireballs observed by the European Network (EN) in 2015 had similar orbital elements and suggested that both asteroids 2015 TX24 and 2005 UR were associated with this stream, and possibly 2005 TF50, arguing that these several hundred metre diameter bodies represented an extension of the population of bodies seen among the observed fireballs. 

\citet{1986A&A...158..259F} first suggested that meteoroids can be trapped by strong mean motion resonances (MMR) with Jupiter. Material trapped in a MMR is prevented from undergoing full nodal precession, explaining concentrations of dust in mean anomaly over long periods of time. \citet{1993MNRAS.264...93A} suggested that this occurred to some Taurids trapped in 7:2 MMR with Jupiter. The expected periodic signature of outbursts was later verified by \citet{1998MNRAS.297...23A}. Their model is successful at explaining enhanced activity in years when the Earth comes within $\Delta_M \in \pm30/40\degr$ of the resonance centre in mean anomaly (Table \ref{table:swarmyears}). \citet{1998MNRAS.297...23A} also published future and past year outburst predictions by his model.

{\catcode`\&=11
\gdef\1997PandSS...45..541S{\citet{1997P&SS...45..541S}}}
{\catcode`\&=11
\gdef\2017AandA...605A..68S{\citet{2017A&A...605A..68S}}}
\begin{table*}
	\label{table:swarmyears}     
	\centering                                 
	\begin{tabular}{cccc}
		\hline  
		\hline  
		Year & $\Delta_M$         & Observations                \\
		\hline 
		1995 & $+29\degr$  & \1997PandSS...45..541S \\
		1998 & $-13\degr$  & \citet{2004Obs...124..277B} \\
		2005 & $+11\degr$  & \citet{2007MNRAS.376..890D,2012LPICo1667.6436S,2017MNRAS.469.2077O} \\
		2008 & $-30\degr$  & \citet{2009JIMO...37...55S} \\
		2012 & $+35\degr$  & \citet{2014Icar..231..356M} \\
		2015 & $-07\degr$  & this work; \2017AandA...605A..68S \citet{2017MNRAS.469.2077O} \\
		2022 & $+17\degr$  & upcoming return \\
		\hline
	\end{tabular}
	\caption{Predicted returns of the Taurid Swarm. Updates for recent years are published at the website \url{https://www.cantab.net/users/davidasher/taurid/swarmyears.html}, accessed May 16, 2017.}
\end{table*}

\citet{2016Icar..266..331J} have identified this shower in 2010 - 2013 CAMS data as \#628 in the IAU Working List of Meteor Showers, and called it the s-Taurids (IAU code \textit{STS}). We note that the "new stream" of \citet{2017A&A...605A..68S} corresponds to the same IAU \#628. Hereafter meteor showers IAU \#2 (codenamed STA) refers to the "regular" Southern Taurids, IAU \#628 (codenamed STS) designates the resonant Southern Taurid branch (s-Taurids), and the Southern Taurid Complex encompass members from both sub-streams.

\citet{2017A&A...605A..68S} outlined a correlation between size and strength: larger bodies among the 2015 Taurids tend to be more fragile. If so, that would imply that metre-sized objects would break at such high altitudes in the Earth's atmosphere that they might be recognised in satellite observations. More recently, \citet{2020P&SS..18204849B} found from a sample of 16 studied Taurid fireballs that the meteoroids $>10$\,cm in size had low tensile strength, less than 0.01\,MPa, and a density less than 1\,g\,cm$^{-3}$. Smaller meteoroids contain a higher fraction of materials up to 0.3\,MPa in strength. 

With an eye on the upcoming 2022 return of the s-Taurids, we present in this paper observations of the enhanced 2015 Taurid fireball activity as observed by the Desert Fireball Network (DFN) in Australia and by the Cameras for Allsky Meteor Surveillance (CAMS) network in California. We investigate changes in the orbital elements along Earth's path, the stream's semi-major axis distribution and the particle size distribution of the stream in order to better understand its relationship to comet 2P/Encke and other Taurid Complex parent bodies. 

\section{Data and methods}

DFN and CAMS survey meteoroid impacts at different sizes ranges: CAMS has the sensitivity to detect large numbers of small millimetre to centimetre size grains, while the DFN takes advantage of a large collecting area to catch centimetre to decimetre scale meteoroids, at the cost of lower sensitivity. When it comes to observing a bright meteor shower like the Taurids, the two systems complement each other well.

\subsection{DFN} \label{sec:dfn_red_methods}

\begin{figure}
	\centering
	\includegraphics[width=1\linewidth]{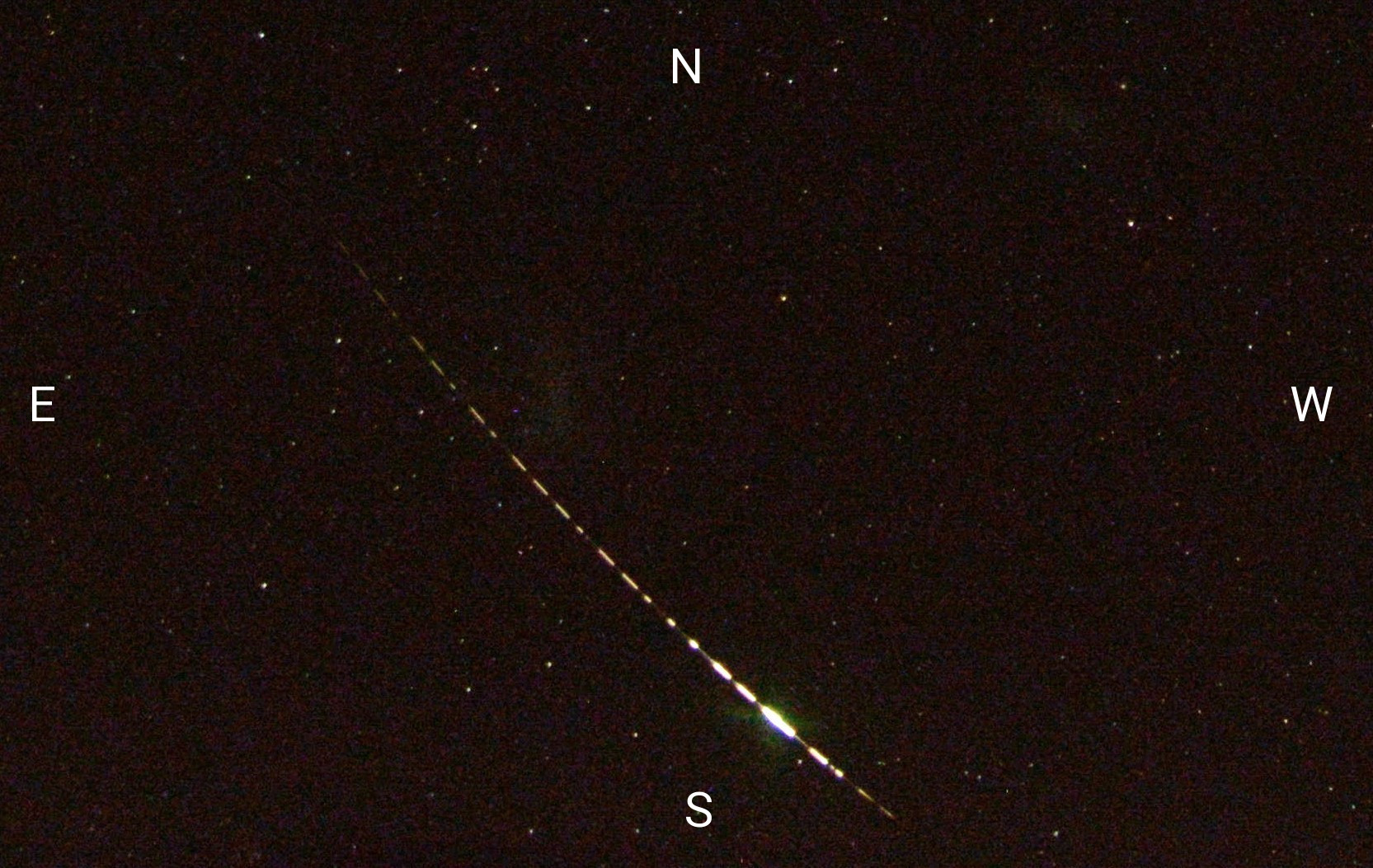}
	\caption{DN151104\_01: a 2.6\,s s-Taurid observed at Hughes siding in the Nullarbor plain, near the Magellanic clouds.	This is a crop of the original all-sky picture. The meteoroid experiences a catastrophic fragmentation at 74\,km altitude, shortly before the meteor faded.}
	\label{fig:taurid_pic}
\end{figure}

In 2015, the Australian Desert Fireball Network covered 1.5 million $km^2$ of sky viewing area, established around $30\degr$ S latitude \citep{2017ExA...tmp...19H}. Each DFN observatory comprises of a high-resolution still imaging system: a 36\,Mpixels digital camera (Nikon D800, D800E, or D810), associated with a Samyang 8\,mm f/3.5 fish-eye lens, taking 25 seconds exposures at 6400 ISO. In 2015 all observatories operated with these settings. The field of view of the cameras is all-sky, except for a crop of 10 degrees on the horizons of long sides of the sensor (usually North and South). The pixel size is 119 seconds of arc in the center, decreasing towards the edges (87 seconds of arc at $5^\circ$ elevation). The cameras are sensitive to stellar magnitude 0.5 for meteors (7.5 for stars), and reliably detect meteors that are brighter than apparent magnitude -1.5 for $>0.9$ second.

Meteor events are automatically detected in the images by the software procedures described by \citet{towner2017detection}. Astrometric measurements are performed in the same manner as described by \citet{2018M&PS...53.2212D}, resulting in measurements precise down to 1-2 minutes of arc. The triangulation of meteor trajectories are performed using a weighted straight line least squares approach, similar to the one described in \citet{1990BAICz..41..391B}. In order to get an appropriate entry velocity for the meteoroid, an extended Kalman smoother is applied to the positional data, throughout the visible bright flight \citep{2015M&PS...50.1423S}. This method also yields statistical uncertainties that encompass both model errors and measurement errors. These results are crucial for initialising orbit determination as orbital parameters, like the semi-major axis and eccentricity, are very sensitive to the errors in initial velocity. The heliocentric orbit of the meteoroids are determined using a backward integration from the start of the visible bright flight. The meteoroid is back-tracked through the upper layers of the atmosphere, and out of the sphere of influence of the Earth to a distance of one Hill sphere \citep{2019M&PS...54.2149J}. Uncertainties on the orbital parameters are computed using a Monte Carlo method based on the uncertainties of the first velocity vector observed.

The DFN data reduction pipeline uses aperture photometry on the fireball track to calculate brightness. Doing photometry on the reference stars used for astrometry yield instrumental zero point of each camera, accounting for extinction and vignetting. The fireball brightness is converted into magnitudes by accounting for the different exposure times: the effective exposure time for stars is typically 11.2\,s (25\,s exposure modulated by the liquid crystal shutter), and 0.06\,s or 0.02\,s for a fireball shutter break (see \citet{2017M&PS...52.1669H} for details on the action of the liquid crystal shutter). Apparent magnitude is converted to absolute (for constant distance of 100 km) after triangulation, using the observation range. The main limitation on this technique is the saturation of the sensor, which typically happens when the fireball exceeds apparent magnitude -6. Blooming of the trail enables brightness measurements out to about -10 magnitude. 

The main use of photometric measurement in the present study is to calculate meteoroid strength and to get a zero-order mass estimate. As detailed by \citet{2016Icar..266...96B}, the peak brightness instant of a fireball is a good indicator of catastrophic fragmentation, and therefore a reasonable proxy for calculating a bulk tensile strength for the entering body. This method is more robust to instrumental bias than the PE criterion introduced by \citet{1976JGR....81.6257C}, and has the advantage of being inferred directly from observable parameters (no mass calculation involved). We therefore use the following relation from \citet{1981MoIzN....Q....B} to calculate tensile strength $S$: 
$ S = \rho_{atm} v^2$, where $\rho_{atm}$ is the density of the atmosphere estimated using the \textit{NRLMSISE-00} atmospheric model \citep{2002JGRA..107.1468P}. $v$ is the velocity at that instant calculated by the Kalman smoother described by \citet{2015M&PS...50.1423S}.
The main limitation of the method comes from the uncertainty on the instant of peak brightness, dominated by the sampling rate (10\,Hz), which translates into 2\,km of altitude for the average Taurid, or a $\simeq1.3$ factor in strength.

Thanks to the continental scale of the network, operational and weather biases are mitigated by the large collecting area and observation time. However a consequence of this is that precisely determining the surveying area probed by the instrument is difficult. While calculating probing area as a function of time may be done accurately and relatively easily when a small number of narrow angle optics are used, such as described in \citet{2016MNRAS.463..441B}. Even at a basic level, this kind of work with all-sky cameras spaced on a continent-scale network, is more tricky, and de-biasing the DFN dataset to get precise fluxes will be the subject of a future paper.

The DFN observatories, combined with the data reduction methods described above, have led already to multiple meteorite recoveries: Creston \citep{2019M&PS...54..699J}, Murrili \citep{2020M&PS...55.2157S}, Dingle Dell \citep{2018M&PS...53.2212D}, as well as 3 more recent (not yet published) recoveries. These successes in precisely pinpointing the location of meteorites are good indications that the data reduction process is free of major systematic issues.

\subsection{CAMS}

The main goal of CAMS is to map the presence of meteor showers of +4 to -5 magnitude meteors throughout the year. In November 2015, CAMS networks had been established in California, Arizona, Florida, the BeNeLux, and New Zealand. Most CAMS networks are on the northern hemisphere, but they experienced a relatively small number of cloudy days that year. CAMS methods are described in detail in \citet{2011Icar..216...40J}. In brief, CAMS utilises a network of analog low-light video cameras, mostly Watec Wat902H2 Ultimate cameras with $30\degr\times20\degr$ field of view each and +5.4 stellar limiting magnitude. Customised software detects the meteors, calibrates the background star field to obtain astrometric positions, and then combines such data from two or more stations to triangulate the meteor trajectory. CAMS yields more than 100,000 meteoroid orbits per year, and has proven to be a very efficient tool for studying meteor showers and linking them to possible parent objects \citep{2016Icar..266..331J, 2016Icar..266..355J, 2016Icar..266..384J}. 

The high detection rate of meteors not assigned to showers by \citep{2016Icar..266..384J}, with geocentric entry speed $<$ 35 km/s from the antihelion source at the same time as the Taurid showers, provides a baseline of sporadic meteor shower activity that can be used to calculate the effective observing time due to weather. The CAMS flux data were de-biasing by assuming a constant sporadic flux during the s-Taurid activity period.  The main CAMS networks are situated at a latitude of $\phi=+37\deg$, where the Southern Taurid radiant is up almost all night in early November and the mean altitude of the Southern Taurid radiant is $hR \simeq 40\deg$. This results in a correction factor of $1/sin(hR) = 1.35$ to get the equivalent zenithal hourly rate. For scaling the distribution we use the flux density of \citet{1985Icar...62..244G}, $f_G = 6.85 \times 10^{-8} >1\,\mbox{g meteoroids m}^{-2}\, \mbox{year}^{-1}$, and the correction factor between interplanetary and top of the atmosphere $S=0.67$ be of \citet{2019JSpRo..56.1531M}. From this, we can calculate the influx of s-Taurid on the Earth as:
$$N(\#628 >1\,\mbox{g impacts on Earth}) = \frac{n}{m} * \frac{1}{\sin(hR)} * f_G * S * 2 * \Delta\lambda_{\odot} * \mbox{X-section}_{\mbox{Earth}} $$

In this formulation $\Delta\lambda_{\odot}$ is the exposure time to the shower as observed by CAMS, and the factor 2 is there to compensate that a surface at top of the atmosphere is effectively twice the collecting area of the model randomly tumbling plate of \citet{2019JSpRo..56.1531M}. In relation to a meteoroid stream, the Earth's exposure to the stream can be effectively represented as a cross-section area defined by its radius.

The reported flux values are limited to a 1\,g threshold mass using the observed magnitude distribution. The count of all sporadic meteors with the 25-30 km/s entry speed of Taurids was assumed to be exponential in shape of this magnitude interval, from which a detection probability function was derived by fitting an exponential slope to the bright-end of the magnitude distribution and then dividing observed counts by the fit-predicted count. The fraction of completeness for magnitudes -1 and up was: P(m) = 1.00, 0.80 $\pm$ 0.02, 0.37 $\pm$ 0.01, 0.093 $\pm$ 0.003, 0.014 $\pm$ 0.002, 0.0011 $\pm$ 0.0002, and $\sim$5e-6. This probability function was then applied to the detected count of shower meteors to derive the magnitude size distribution of different Taurid Complex component showers.

\begin{figure*}
	\centering
	\includegraphics[width=1\linewidth]{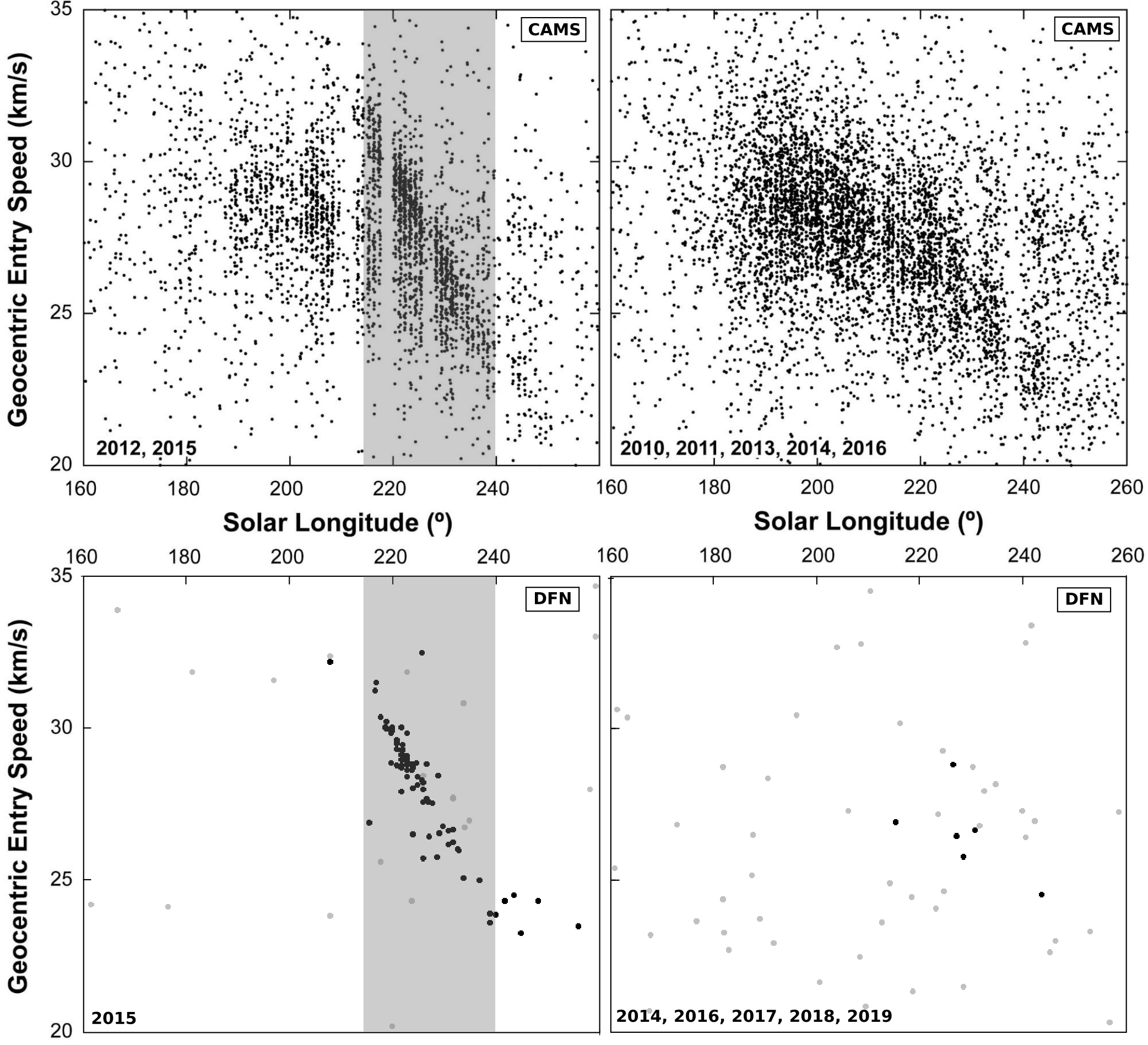}
	\caption{Detected meteors geocentric entry speed as a function of solar longitude.
		STS activity years (left) are separated from other observation years (right).
		In DFN data, fireballs identified as Southern Taurids are marked in black against grey background fireballs. While the STS component of the Southern Taurids, recognised by its strong date-speed linear relationship, is visible in the meteor data (CAMS, top left), it is even more obvious at fireball sizes (DFN, bottom left).}
	\label{fig:Vg_vs_Sol_long}
\end{figure*}

\section{Results}
The 73 measured DFN orbits from the Southern Taurid Complex in 2015 are provided as an appendix to this manuscript. The CAMS-derived Southern Taurid Complex orbits were released as part of the 2013 – 2016 CAMS data release and can be accessed via the project website (http://cams.seti.org) and via the Meteor Data Center. In 2015, CAMS detected N = 10,942 Southern Taurids (N = 177 $>1$\,g) between $\lambda_{\odot} \in [213.19$-$234.25]\degr$. In the same period, 131,230 (Nspo =  1193 $>1$\,g) sporadic meteors were recorded. 

\subsection{Comparison of s-Taurids with regular Southern and Northern Taurids}

The s-Taurid shower stands out well from other Southern Taurid complex meteors by their geocentric speed. Fig. \ref{fig:Vg_vs_Sol_long} plots the geocentric speed and time (solar longitude) of all meteors associated with the Southern Taurid complex in both the CAMS (top) and DFN (bottom) datasets. Vertical white bands are due to cloudy weather with less than complete coverage. The data are split in two groups: the outburst years of 2012 and 2015 (Table \ref{table:parentbody}), and the no-outburst years of 2010, 2011, 2013, 2014, and 2016. The 2012 encounter with the s-Taurids is $\Delta_M = 35\degr$ from the centre of the resonance according to the model of Table 1. Hence, the weak detection of the s-Taurids in 2012 by CAMS implies an extend of this component until at least mean anomaly $35\degr$, in agreement. 

The outburst years show a component that produces a narrow range of geocentric entry speed at any given solar longitude, with a strong change in the speed as a function of time. This component is only weakly present in non-outburst years (Fig. \ref{fig:Vg_vs_Sol_long}). This component was earlier identified as shower \#628 , the s-Taurids (IAU code STS). The period of activity for this component is $\lambda_{\odot} \in [213, 234]\degr$. 

The presence of this STS component is also evident in the 2015 DFN data (Fig. \ref{fig:Vg_vs_Sol_long}), despite a lower number of orbits, as the STS stream largely dominates the Southern Taurid activity at fireball sizes. The change in velocity with solar longitude translates into a strong increase of perihelion distance with increasing solar longitude and a decreasing eccentricity. The semi-major axis and inclination of the orbits remain nearly constant, as does the longitude of perihelion. 

Fig. \ref{fig:cams_rates} shows the de-biased STS rates for CAMS, along with that of the remaining STA and NTA streams. The rates are normalised to that of all sporadic meteors with speeds $< 35\,\mbox{km s}^{-1}$. This ensures that the total sporadic count reflects the observing conditions during that part of the night when the antihelion source is best observed.The sporadic apex and Toroidal sources have been removed from the count. The 2015 STS count was compared to the sporadic meteor rate in 2015 only. The multi-year de-biased distribution produced better defined shower activity profiles than early results in \citet{2016Icar..266..355J}. The shower components identified in \citet{2016Icar..266..384J} are still present. The STA and NTA shower profiles are different, an indication that the nodal line of individual meteoroid orbits did not fully rotate, as earlier pointed out.

\subsection{Size frequency distribution of the \#628 STS stream}\label{sec:sfd}

\begin{figure}
	\centering
	\includegraphics[width=0.5\linewidth]{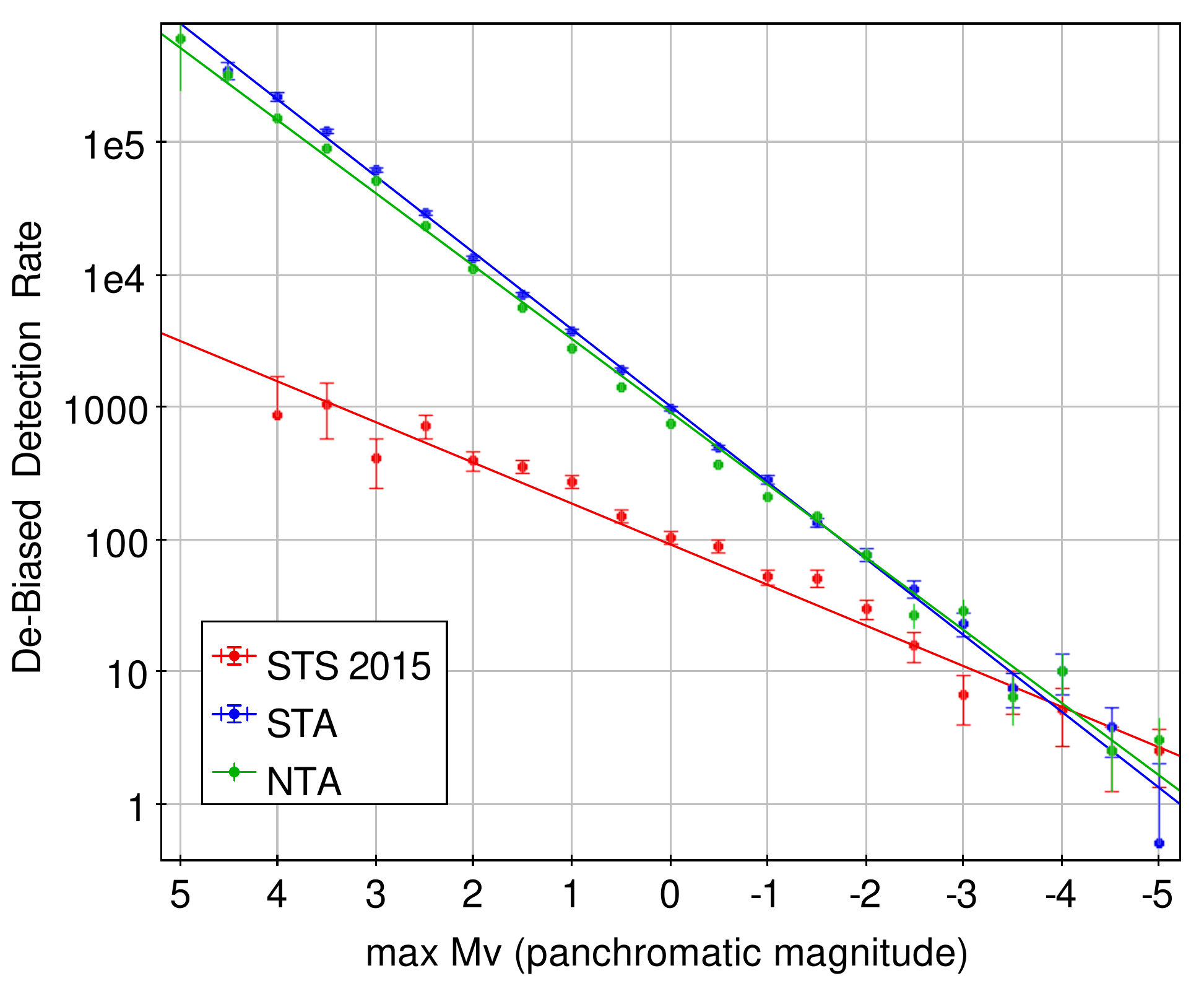}
	\caption{Peak magnitude frequency distribution for Southern Taurids substreams \#2 STA, \#17 NTA and \#628 STS. Resonant Taurids (STS branch) are generally larger than regular Southern Taurids (STA). 
	}
	\label{fig:mv_distribution}
\end{figure}

Fig. \ref{fig:mv_distribution} shows the distribution of peak magnitudes in 0.5 magnitude intervals for CAMS-detected NTA, STA and STS meteors.  The STS population is significantly more skewed towards brighter meteors. The cumulative mass-frequency distribution as observed by CAMS (Fig. \ref{fig:SFD}) can be expressed as $N(>m) = a m^{-b}$ with $a=1.42\times 10^{5}$ and $b=0.94$ when expressed in grams. This is representative of the total influx of s-Taurids during the 2015 encounter.

\begin{figure}
	\centering
	\includegraphics[width=0.7\linewidth]{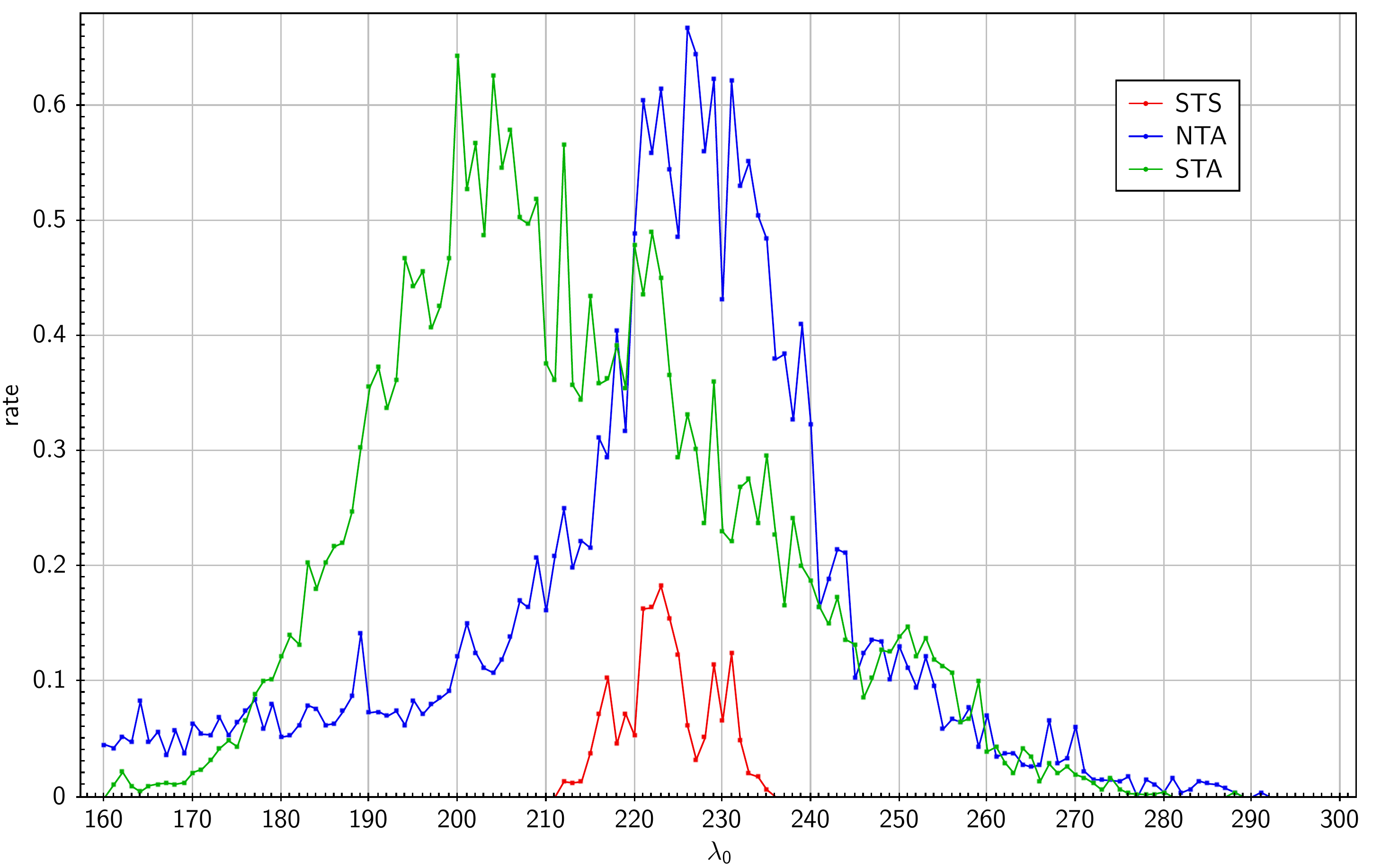}
	\caption{De-biased CAMS shower rates relative to that of sporadic meteors $<$35 km/s for shower 628 (STS), the Southern Taurids (STA) and the Northern Taurids (NTA), as a function of solar longitude.}
	\label{fig:cams_rates}
\end{figure}

The annual Southern and Northern Taurids have differential magnitude indices of $\chi \simeq 3.0$ (Fig. \ref{fig:cams_rates}), this is close to the typical value for JFC material ($\chi = 3.29 \pm 0.09$, as determined by \citet{2016Icar..266..384J}). On the other hand s-Taurids have a much shallower distribution with $\chi = \sim 2.0$ (assuming a sporadic $\chi = 3.4$). This gives a differential mass index for the observable stream of $s=1.75$ ($s = 1 + 2.5 \log(\chi)$).\ 

The observed size distribution confirms that the s-Taurids are relatively rich in large meteoroids. Indeed, a study by \citet{2011MNRAS.414.1059S} on radar meteor observed by the Canadian Meteor Orbit Radar (CMOR) in 2005 (typical observed mass of $10^{-7}$\,kg, which roughly corresponds to optical magnitude +7), failed to identify the 7:2 resonance from regular Southern Taurids. They discuss that this is partly due to the poor constraints the radar observations put on the velocities (and therefore the semi-major axes), so it is not possible to distinguish STSs from STAs dynamically.Therefore unless the STS outburst is strong enough to significantly skew the overall Southern Taurids rates, it is not detectable. \citet{2011MNRAS.414.1059S} do no provide an upper constraint on the STS/STA activity, but even without hard numbers this analysis confirms the trend shown in Fig. \ref{fig:mv_distribution}: the STA branch dominates the STS branch at the low mass end ($Mv_{max} > -4$).

\begin{figure}
	\centering
	\includegraphics[width=0.8\linewidth]{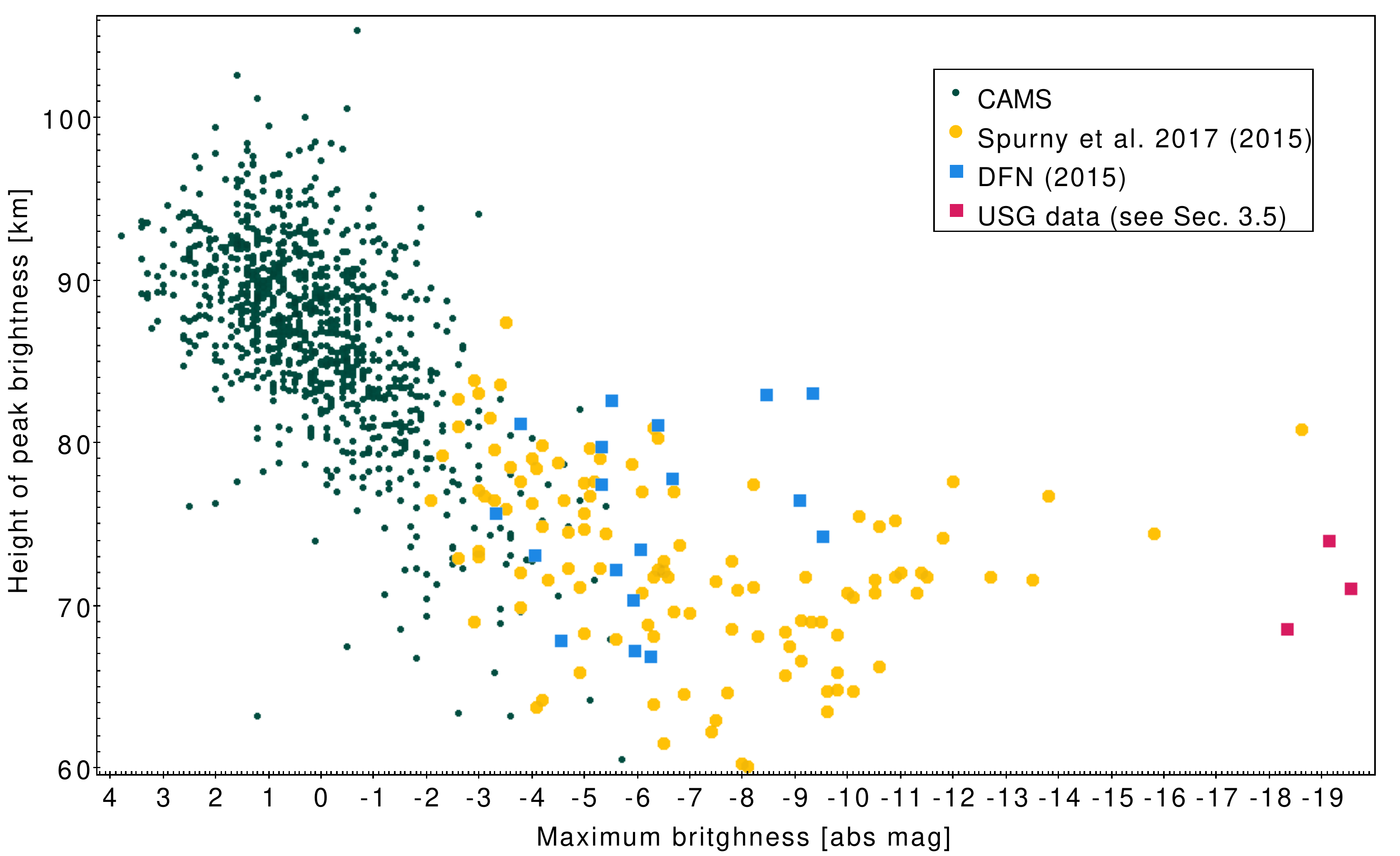}
	\caption{s-Taurids height of peak brightness  as a function of brightness. DFN magnitudes may be slightly underestimated because of saturation issues.}
	\label{fig:heights}
\end{figure}

Beyond CAMS data, towards fireball sizes, flux density data for this stream becomes more scarce. The DFN and EN fireball networks do not yet have time-area de-biasing information to calculate flux densities. The masses reported by \citep{2017A&A...605A..68S} give an idea of the slope of the distribution at magnitudes below -9, where the EN sample appears to be complete (Figure \ref{fig:SFD}). Although it is apparently more shallow than the slope calculated by CAMS, the numbers are so small that the extrapoplated flux agrees within 2 error bars (2 sigma). Even at these bright magnitudes there may still be an effect of the brightness-dependent variation in effective covering/reporting: the brighter the bolide, the further it can get detected and studied.

\subsection{Strength of the meteoroids in the \#628 STS stream}

Figure \ref{fig:heights} shows the altitude of peak brightness as a function of peak meteor magnitude. The result shows that larger meteoroids penetrate deeper in Earth's atmosphere before reaching peak brightness, as expected. Among CAMS-detected visual meteors, that is a fairly continuous trend.
However, at the larger fireball sizes, meteoroid penetration enter a strength dominated regime. The transition into the strength-dominated regime occurs at about -7 magnitude. In the strength-dominated regime, the altitude of peak brightness is independent of mass, although there is a weak trend that the largest meteoroids are weaker than the smaller meteoroids near this transition. All s-Taurids experience their peak brightness above 60\,km altitude (Fig. \ref{fig:heights}), and do not survive below 50 km altitude.
The deepest penetrating fireball had a minimum height of 54.5\,km, while the average end height of the DFN fireballs was 67.2\,km (see suppl. mat. table). This is consistent with the results of \citet{2017A&A...605A..68S}.

\begin{figure}
	\centering
	\includegraphics[width=0.5\linewidth]{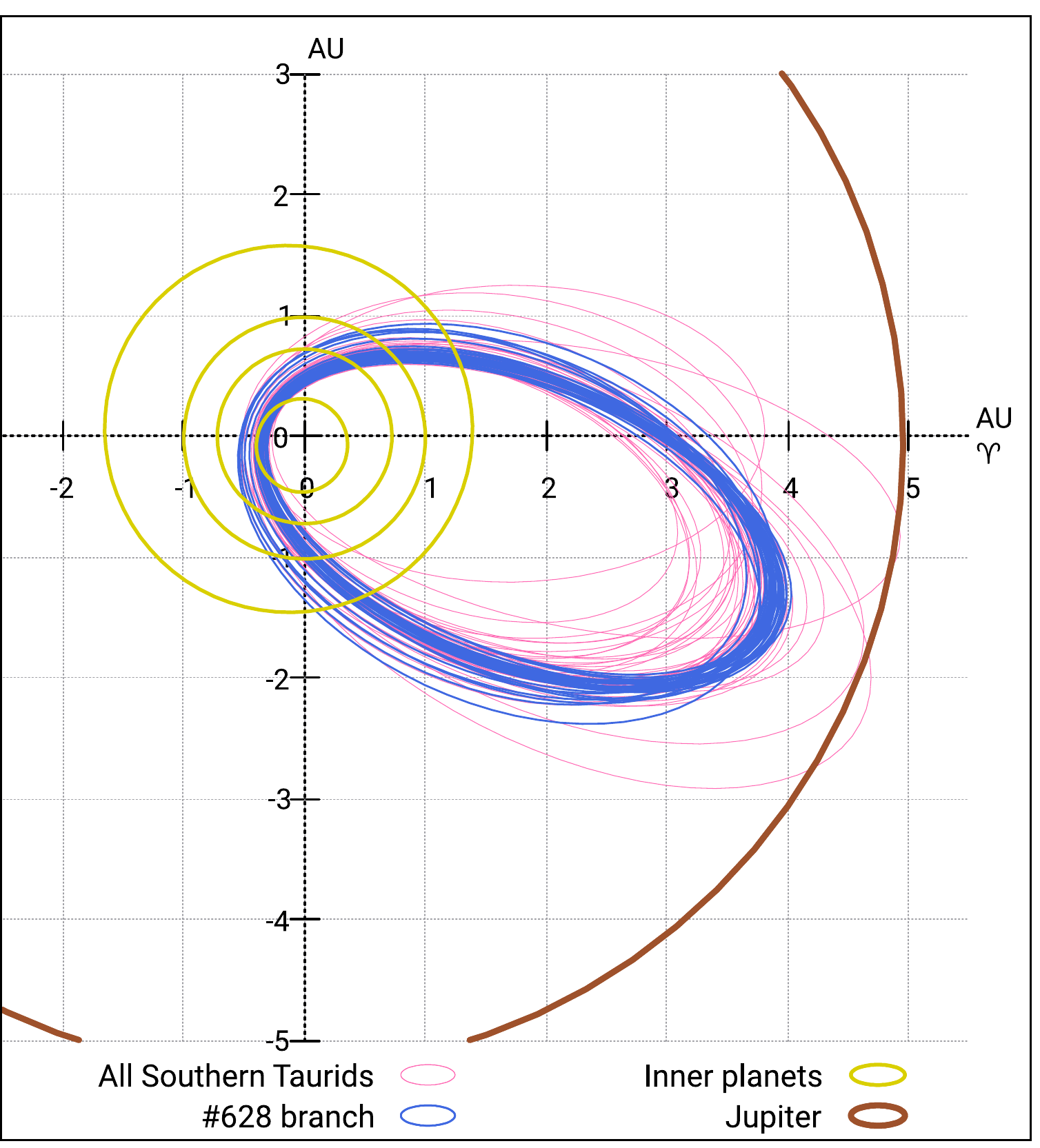}
	\caption{Ecliptic orbit plot of all Southern Taurids observed by the DFN in 2015 (pink), and the \#628 branch (blue). }
	\label{fig:sta_orb_plot}
\end{figure}

\begin{figure}
	\centering
	\includegraphics[width=0.5\linewidth]{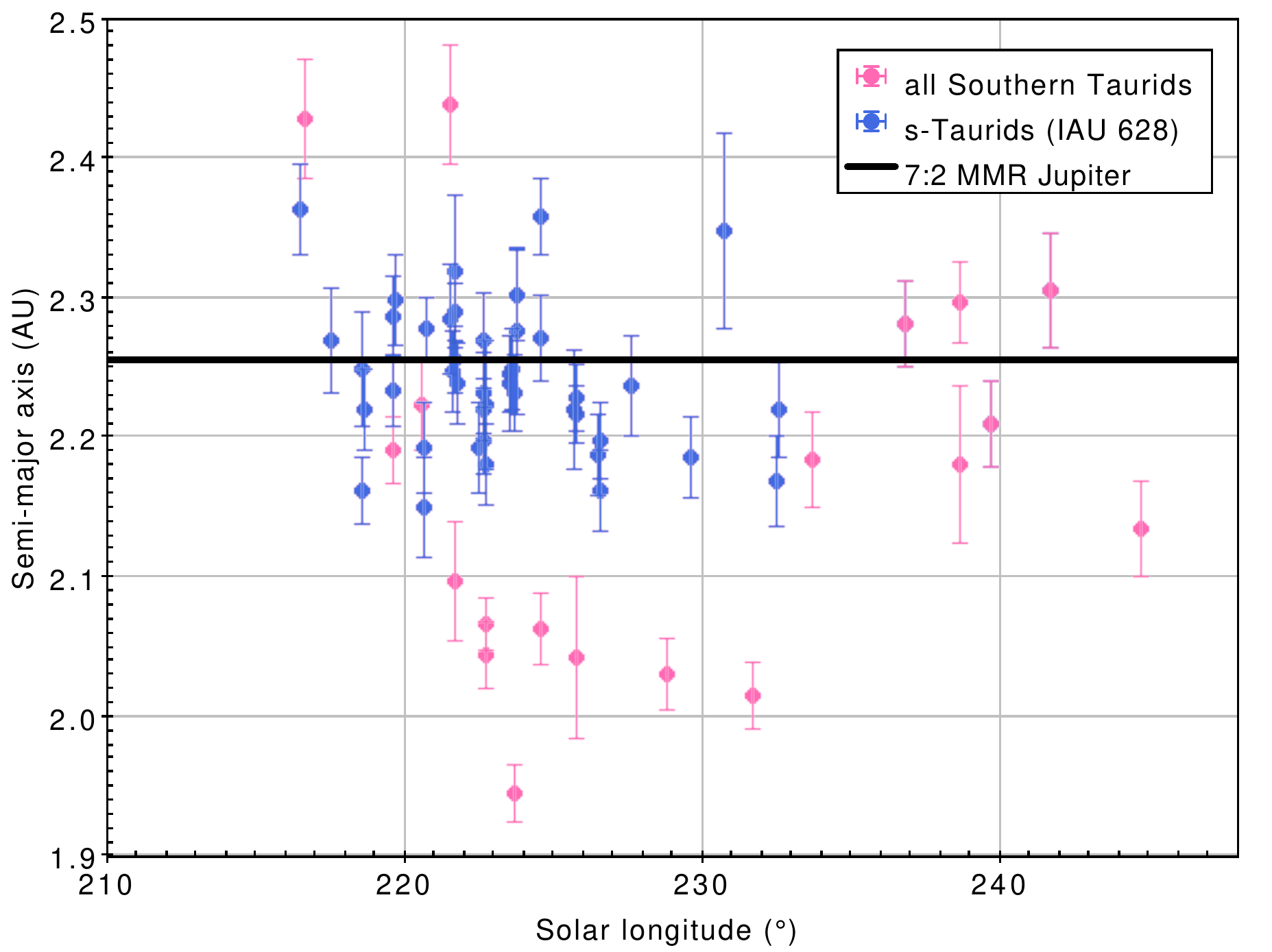}
	\caption{Semi-major axis measurements (with $1\sigma$ uncertainties error bars) of all Southern Taurid fireballs observed by the DFN 2015. Most are significantly higher than typical Southern Taurids, compatible with the a 7:2 mean-motion resonance with Jupiter (centered on 2.256\,AU).}
	\label{fig:SMA_vs_Sol_long_DFN}
\end{figure}

\subsection{Semi-major axis of the fireball orbits}
Figure \ref{fig:sta_orb_plot} is an ecliptic orbit plot of all Southern Taurids observed by the DFN in 2015. The Figure shows that most observed Taurid fireballs clustered in a tight stream with constant longitude of perihelion (all ellipses pointed in the same direction). These s-Taurids are shown in blue. This stream is highly stratified: they form a series of more ore less concentric ellipses for fireballs detected at different solar longitude along the Earth's orbit. Unlike most meteoroid streams, this stream appears to be more narrower at aphelion than at perihelion. The dispersion of perihelion distances (q) of the blue orbits is 8.9 percent (one standard deviation), while the aphelion distance (Q) is dispersed by only 2.4 percent, and the semi-major axis (a) by 2.1 percent. a and Q are tightly correlated, a and q are not. The stream approaches the orbit of Jupiter near aphelion, suggesting that the secular perturbations responsible for the observed dispersion are strongest near aphelion, not near perihelion. 

Figure \ref{fig:SMA_vs_Sol_long_DFN} plots the semi-major axis of the 2015 Taurid fireballs measured in the DFN network as a function of solar longitude (time in the year). The DFN-derived orbits during solar longitudes 217.5 and 227.5 show a clear concentration of semi-major axis values around the mean semi-major axis a = $2.234 \pm 0.007$ AU with a standard dispersion of 0.041 AU. The mean calculated error in the semi-major axis values is $0.034\pm0.014$\,AU, in good agreement with the observed dispersion if all these orbits have exactly the same semi-major axis of a = 2.2563 AU corresponding to the 7:2 mean-motion resonance with Jupiter (dashed line). This result confirms earlier conclusions from EN fireball observations reported by \citet{2017A&A...605A..68S} that the meteoroids appear to be trapped in this resonance, and demonstrates the accuracy of the semi-major axis calculations. Most of the observed dispersion in semi-major axis is due to measurement error. However, there is a small systematic error of $-0.022\pm0.007$\,AU.

The fireballs shown in pink are outside the main dispersion. Most have slightly smaller semi-major axis, while three orbits just reach the orbit of Jupiter. 52 of the observed trajectories have a semi-major axis within 2 sigma from the resonant value. 21 have not. Surprisingly, the once that have not scatter around the time of the outburst over 216.5 to 231.7 degrees solar longitude (CAMS has 212 to 235 degrees range, see below), while the once that are within 2 sigma from the resonant value scatter over the whole observing interval of 207.7 to 255.9 solar longitude. Hence, whether or not the trajectories scatter around the resonant value is not a mark of the s-Taurids, it is a mark of the Southern Taurids as a whole. 

Some of the pink orbits may represent measurement errors in the initial velocity determination. Or they are a more perturbed population of meteoroids that possibly already experienced some changes in semi-major axis due to encounters with the terrestrial planets. If so, they are likely part of the background Southern Taurids.

\begin{figure}
	\centering
	\includegraphics[width=1\linewidth]{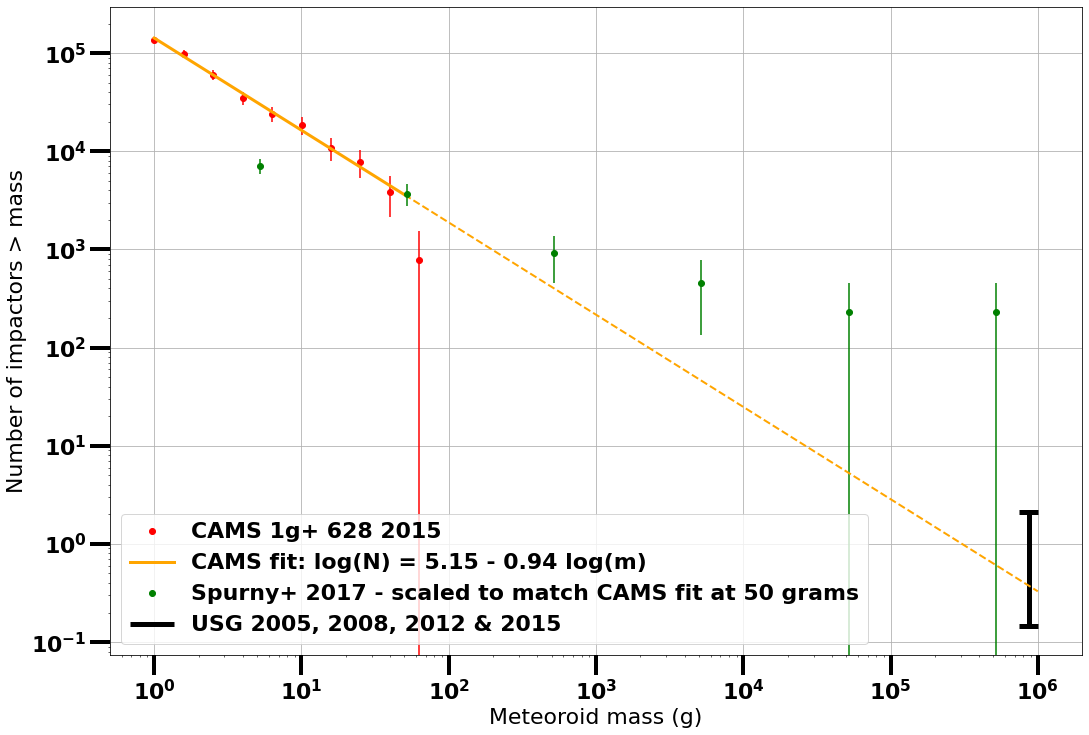}
	\caption{Cumulative mass-frequency distribution. CAMS $>1$\,g meteors left is fitted based on the 2015 data (see Sec. \ref{sec:sfd}). The s-Taurid 2015 data of \citet{2017A&A...605A..68S} is presented in green, with the event frequency scaled to match CAMS' at 50 grams. Note that the different slope for this data may not be real and could come from a brightness-dependent variation in effective covering area. The right-hand side estimate comes from 3 likely large s-Taurids observed by the USG sensors in 2005 and 2015.}
	\label{fig:SFD}
\end{figure}

\subsection{The metre-size population}\label{sec:metre-size_pop}

\citet{2017A&A...605A..68S} first pointed out that the stream contains metre-sized asteroids. Here, we will attempt to quantify this. The US government (USG) satellite sensors detect m-scale impactors in Earth’s atmosphere over the entire planet as a collecting area, or if a meteoroid stream is involved approaching from a given direction that collection area is the projected Earth surface to that stream. These data are reported online on the NASA JPL fireball website\footnote{\url{https://cneos.jpl.nasa.gov/fireballs/}, accessed May 16, 2017}. Even with this large a collection area, the chance of detecting impacts is small. Figure \ref{fig:SFD} shows the cumulative mass-frequency distribution from CAMS and DFN data. The USG detected bolides are about -18 magnitude and brighter \citep{1995hdca.book.....G}. That leaves a large gap without de-biased data. 

Extrapolating the mass-frequency distribution established in Sec. \ref{sec:sfd}, we predict 0.37 meteoroid $>940$ kg  impacting the Earth during the 2015 s-Taurid episode. Indeed, the USG satellites detected no unusual number of fireballs in late October and early November 2015. One bolide was detected on 2015-10-31, one on 2015-11-02 and one on 2015-11-13. No velocity components are reported, so we do not know the radiant or speed of these bolides.

\begin{table*}
	\caption{14 high altitude (>60 km) meteoroid airbursts observed by the USGS. No velocity information was provided for the events presented here. Highlighted rows show events that fall within the STS activity period. The 4 events at the bottom are excluded for statistical significance reasons (see in the text). Size is calculated from the energy, assuming the velocity is equal to the mean STS velocity observed by the DFN at the same given solar longitude (see Suppl. Mat. Table), and the 1600\,$\mbox{kg m}^{-3}$ bulk density estimated by \citep{2009A&A...495..353B}.}  
	\label{table:usgs_hi_alt_airbursts}      %
	\centering                                      %
	\begin{tabular}{ccccccc}
		\hline  
		\hline  
		Peak brightness time & $\lambda_{\odot}$& Latitude & Longitude & Altitude & Total Impact Energy & Size \\
		ISO UTC & $\degr$ & $\degr$ (N+) & $\degr$ (E+) & km & kilotons TNT & m \\
		\hline  
		\rowcolor{Gray} 2015-10-31 11:34:30 & 217.51 & 9.0 & -138.0 & 71.0  & 0.29 & 1.5 \\
		2015-06-10 17:43:03 & 79.32 & -11.5 & -161.9 & 61.1  & 1.0 & \\
		2013-08-12 18:08:02 & 140.00 & -34.4 & 118.2 & 66.6 & 0.15 & \\
		2012-02-12 05:25:52 & 322.72 & -31.7 & 54.9 & 61.0 & 0.41 & \\
		2011-01-21 15:11:43 & 301.07 & 18.9 & -44.6 & 61.0  & 0.23 & \\
		2010-12-09 02:54:07 & 256.76 & -54.5 & -169.7 & 66.0  & 0.2 & \\
		2005-12-24 15:30:26 & 272.85 & -54.0 & 17.3 & 66.0  & 0.51 & \\
		\rowcolor{Gray} 2005-11-02 07:04:32 & 219.89 & 33.9 & -154.9 & 68.5  & 0.11 & 1.1  \\
		\rowcolor{Gray} 2005-11-02 05:16:47 & 219.81 & 22.9 & -123.8 & 74.0 &  0.21 & 1.3  \\
		2005-04-06 01:30:24 & 16.28 & -42.7 & 154.6 & 70.0  & 0.1 & \\
		\hline 
		*2011-08-04 07:25:57 & 131.44 & -40.7 & -86.7 & 63.0  & 0.098 & \\
		2004-01-02 04:27:59 & 281.05 & -28.2 & 3.2 & 63.0  & 0.39 & \\
		\rowcolor{Gray} 1999-06-25 06:27:41 & 93.30 & 50.0 & 121.0 & 69.0  & 0.37 & \\
		1999-01-02 18:25:51 & 281.93 & 47.0 & 103.0 & 65.0  & 0.12 & \\
		\hline  
	\end{tabular}
\end{table*}

However, there are data from multiple years of observation now. If any Taurids are among these impactors, we expect their penetration depth to be relatively shallow based on the size-dependence shown in Figure \ref{fig:heights}.The observed cm-dm sized meteoroids by CAMS, DFN, and EN show no significant decrease in peak height as sizes get larger.

The USG data only include the altitude of peak brightness. We started by filtering the USG dataset by height of maximum brightness $>$60km as a first pass to identify weak cometary impacts, as STS observed by the DFN break up $>66$\,km (Fig. \ref{fig:heights}). We note that the stated peak brightness altitude from the sensors is generally reliable to about $\pm$ 4 km, as shown by \citet{2016Icar..266...96B}, and that these altitudes are reported for most events from the beginning of 2005 onwards. Detections are made at night, but also in daytime. As mentioned by \citet{2019MNRAS.483.5166D}, the typical energy report limit is 0.1\,kT TNT, therefore we exclude event \textit{2011-08-04 07:25:57} (0.098\,kT reported yield) from our analysis for detection significance issues. We are left with 10 significant events that fit the height criterion (Table \ref{table:usgs_hi_alt_airbursts}). However, there are no velocity vectors reported for these events, so there is no direct dynamic link between any of these and the Taurid complex. 

Nevertheless, 3 out of these 10 very weak meteoroids fall in the STS activity period, and even more remarkable they happen in 2005 and 2015, two years during which strong STS activity has been reported and are predicted by the model of \citet{1998MNRAS.297...23A} (Table \ref{table:swarmyears}). In 2005, two events occurred in short succession. All three events suggest the largest fragments in this stream are at solar longitude $[213, 234]\degr$. This combined with data in Fig. \ref{fig:cams_rates}, these USG bolides occurred during the STS activity period. We confirm that these 3 events happened while the Southern Taurid radiant was above the local horizon. 

\begin{table*}
	\caption{Poisson test on the significance of the impact rate increase of metre-sized weak material (peak brightness $>60\,km$) hitting the Earth observed by the USGS during STS outburst episodes ($\lambda_{\odot} \in [213, 234]\degr$). Ranges given are at $2\sigma$ confidence. The influx increase factor during an STS outburst is $[2.1, 46]\times$.}              %
	\label{table:poissontestUSGS}      %
	\centering                                      %
	\begin{tabular}{c c c c c}          %
		\hline                     %
		population & surveyed years &  observed events & $\lambda_{\odot}$ integrated ($\degr$) &  rate ($Earth^{-1}\lambda_{\odot}^{-1}$) \\ %
		\hline      
		weak impactors population & [2005-2016]   & 10  & 3960 & [0.001 - 0.005] \\ %
		probable STS & 2005, 2008, 2012, 2015  & 3 & 84  & [0.007 - 0.1]   \\ %
		\hline
	\end{tabular}
\end{table*}

Can these bolides signify a detection of the s-Taurids? Considering the very low number of events observed, we need to build a statistical test to assess the significance of this apparent rate increase during a swarm episode. Let us test the hypothesis $H_1 $:  "\textit{An airburst from weak material (main explosion $>60$ km) is more likely to happen during a STS activity period}" against the null hypothesis $H_0 $: "\textit{No increase in the rate of impacts from weak bodies can be observed during a STS activity episode}". We define a STS swarm episode as a period that happens on a year predicted by the model of \citet{1998MNRAS.297...23A}, within the interval where the USG sensors have consistently observed airbursts heights (2005, 2008, 2012, 2015), and within the activity period observed by CAMS (solar longitude $\in [213, 234]\degr$). We use the \textit{rateratio.test} \textit{R} package \footnote{\url{https://cran.r-project.org/package=rateratio.test}}, that implements the methods described in \citet{RJ-2010-008} to carry out the statistical test. At 95\% confidence, the background weak metre scale impact rate is $[0.001, 0.005]$, compared to $[0.007, 0.1]\,Earth^{-1}\lambda_{\odot}^{-1}$ when $\lambda_{\odot} \in [213, 234]\degr$, which corresponds to a weak impactors influx increase of $[2.5, 55]\times$ (see Table. \ref{table:poissontestUSGS} for full test test data and results). The small $p-value = 0.004$ shows strong evidence against the null hypothesis (at 95\% confidence). 

Although we cannot definitely link any individual events with the Taurids, the apparent rate increase in metre-scale weak impactors during the STS outburst activity is statistically significant, and we can say that during an STS outburst episode the Earth is more likely to get impacted by a metre-scale STS than a sporadic meteoroid of the same size. Over an s-Taurid episode, the number of $>0.1$\,kT TNT of s-Taurid impactor is $[0.15, 2.1]$ (Fig. \ref{fig:SFD}). 

Adding this detection rate to the overall picture shows that the meteoroid size distribution does not change down to metre size scale (Fig. \ref{fig:SFD}). This result implies that the \#628 STS stream contains some of the largest meteoroids known to any cometary meteor shower. 

Finally, it is possible that the USG satellite detections do not represent all large bolides in Earth's atmosphere.The European Network detected one superbolide during the 2015 outburst called \textit{EN311015\_180520} \citep{2017A&A...605A..68S}.
The observation of this 1300\,kg bolide over a superbolide coverage area of roughly $0.1\%$ Earth is statistically unlikely, but not impossible (Fig. \ref{fig:SFD}). The \textit{EN311015\_180520} superbolide (0.17 kT TNT total kinetic energy) should have been well within the detection range of the USG sensors, but this bolide was not reported. A single mis-match is not indicative of a particular issue as the USG sensors are known to have had as low as 70-80\% Earth coverage in the past \citep{2002Natur.420..294B}. The USG sensors s-Taurid flux could have been underestimated by a factor of 2 considering both 2008 and 2012 were taken into account for the USG clear-sky calculations (Sec. \ref{sec:metre-size_pop}), as these 2 years were quite distant from the resonance centre. It could mean that the metre-scale s-Taurid is a factor of $~2$ larger than what we calculated here. That would still be consistent with a constant mass distribution index from cm to metre scale (Fig. \ref{fig:SFD}).

\section{Discussion}

\subsection{Nature and origin of the s-Taurids}
The low magnitude distribution index ($\chi = 2.0$) and small differential mass index ($s=1.75$) is atypical of most JFC showers. Such values point to a collisionally relaxed population, with equal combined cross-section area in each magnitude bin (s = 1.67), or a collisional cascade where each meteoroid is broken by a mass just big enough to do so (s = 1.83). This points to a very gentle collision process such as would be experienced during a cometary breakup of relatively strong material with intrinsic low-s size distribution. High velocity collisions in the interstellar medium or other aging processes like grain charging and thermal cycling did not significantly affect the size distribution index of this population of meteoroids. That means that the stream is relatively young and was born from relatively strong cometary material. Weak cometary material breaks into very steep distributions towards small particles \citep{jenniskens2006meteor}.

The unusual large number of shower fireballs detected by DFN and EN points to the presence of large meteoroids in this stream. Continuation of the particle size distribution to larger sizes implies a population of metre-sized objects in this stream that appears to have been detected in USG satellite observations. The presence of such large bodies in the stream is consistent with the material being relatively strong for cometary material in general. This stream is currently a major contributor to the overall population of large weak meteoroids. If the USG statistics are representative of larger cometary impact hazards, the s-Taurids dominate the flux when they are active, increasing the impact risk by $[2.5,55]\times$ ($2\sigma$), and overall they represent a significant fraction of all large cometary impactors, even though their activity period is only a few weeks every couple of years!

Our measured DFN fireball mean semi-major axis of $2.234 \pm 0.007$ AU is 0.019 AU higher than the semi-major axis of 2P/Encke, but 0.022 AU lower than the a = 2.2563 AU corresponding to the 7:2 mean-motion resonance with Jupiter. It is also 0.031 AU lower than the current semi-major axis of 2015 TX24.
Natural oscillations of semi-major axis are about 0.03 AU in this part of the asteroid belt \citep{1998AJ....116.3029N}, so the motion of all objects will be affected by the mean-motion resonance. If the breakup happened in the resonance with small relative ejection velocities, is is likely the resonance prevented the dispersion of the dust by avoiding Jupiter's presence at aphelion when the dust was there.

\begin{table*}
	\caption{Proposed parent bodies for the s-Taurids.}              %
	\label{table:parentbody}      %
	\centering                                      %
	\begin{tabular}{c c c c c c c c c}          %
		\hline                     %
		object & epoch &  $a$ & $e$ & $q$ & $i$ & long. ascen. node & arg. peri. & long. peri. \\ %
		 & TDB &  AU &  & AU & $\degr$ & $\degr$ & $\degr$ & $\degr$ \\
		\hline      
		2P/Encke & 2015-08-04.0 & 2.2152 & 0.8483 & 0.3360 & 11.781 & 334.568 & 186.547 &  161.115 \\ %
		2003 WP21 & 2020-12-17.0 & 2.2620 & 0.7849 & 0.4866 & 4.295 & 37.654 & 124.030 & 161.390 \\
		2004 TG10 & 2020-12-17.0 & 2.2334 & 0.8620 & 0.3083 & 4.183 & 205.073 & 317.381 & 162.454 \\
		2005 UR & 2005-10-26.0 & 2.2492 & 0.8818 & 0.2660 & 6.935 & 20.030 & 140.477 & 160.507 \\
		2005 TF50 & 2020-12-17.0 & 2.2730 & 0.8692 & 0.2972 & 10.725 & 0.564 & 159.962 & 160.526 \\
		2015 TX24 & 2020-12-17.0 & 2.2647 & 0.8724 & 0.2890 & 6.049 & 32.827 & 127.151 & 159.978  \\ %
		\rowcolor{Gray} s-Taurids (DFN) & 2015-11-05.1 & 2.237 & 0.847 & 0.358 & 5.45 & 42.72 & 115.70 & 158.31 \\
		\hline
	\end{tabular}
\end{table*}

\citet{2017A&A...605A..68S} argued that the s-Taurids are the product of fragmentation event (of 2004 TG10?) at about 3.6 AU distance from the Sun, where the fireball orbits have lowest dispersion, with relatively high ejection speeds of 1.5 km/s. The observed asteroids 2005 TF50, 2015 TX24 and 2005 UR (and 2004 TG10) were created in that event and now have evolved along the rotation of the nodal line, now located 2000, 2300 and 2400 yr in rotation behind 2004 TG10. Material not trapped in the 7:2 mean-motion resonance has since been lost.

\begin{figure}
	\centering
	\includegraphics[width=0.9\linewidth]{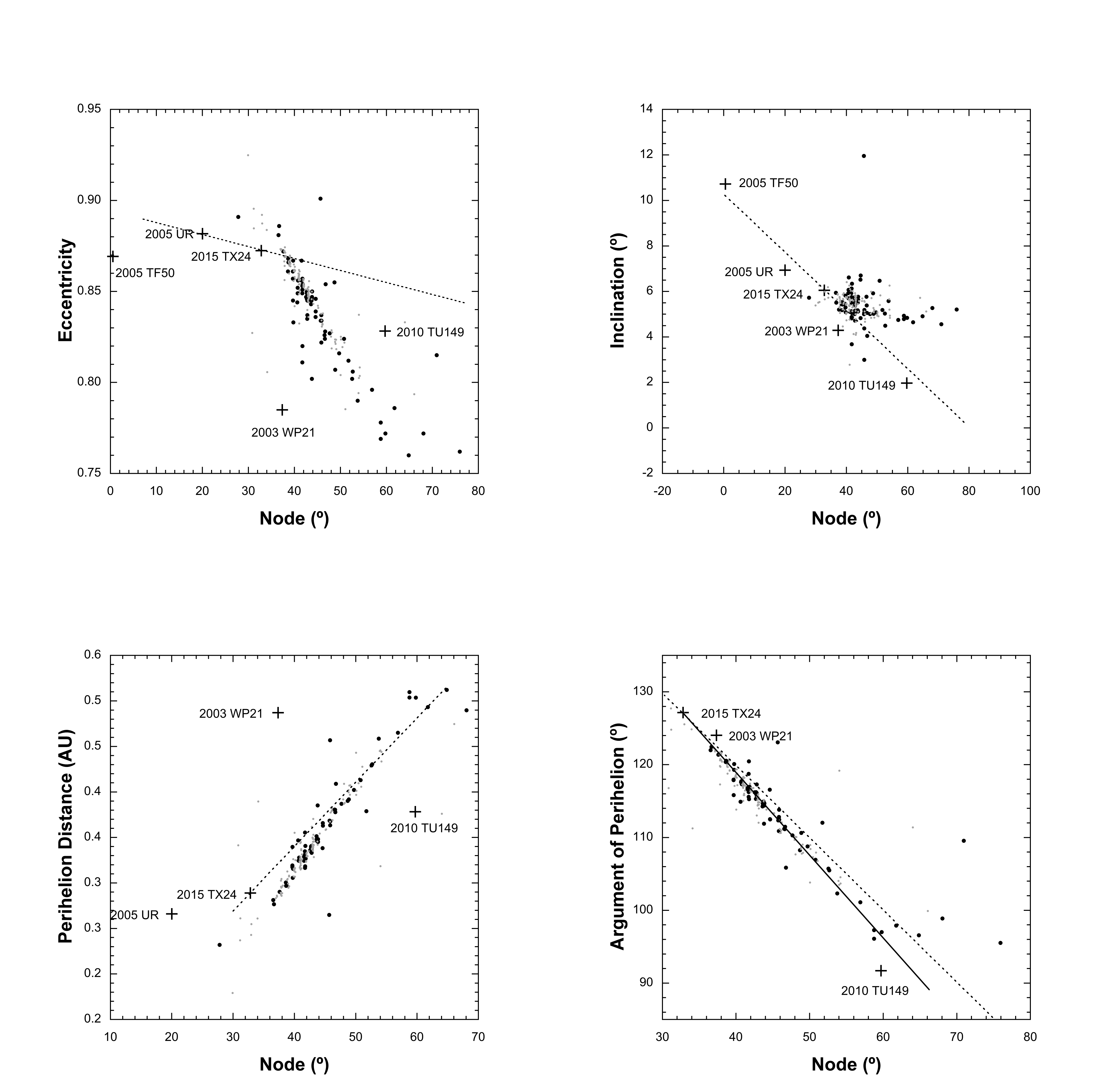}
	\caption{Orbital elements of \#628 meteoroids observed by DFN (black dots) and EN (grey dots, \citet{2017A&A...605A..68S}), put in context with the discussed possible parent bodies for the stream (Table \ref{table:parentbody}).}
	\label{fig:orb_elements}
\end{figure}

That high ejection speed of 1.5 km/s is contrary to the gentle breakup conditions implied by the particle size distribution we measured in this paper. It would also suggest that much of the material was ejected into orbits outside of the mean-motion resonance and material would still be dispersing rather quickly along mean anomaly over time. Instead, we see that dust is confined in a narrow range of mean anomaly. The best handle on age comes from the work by \citet{1952HelOB..41....3W} and \citet{2016MNRAS.461..674O} who found that individual fireball orbits originated from a common orbit about 1,500 years ago. However, if the mean-motion resonance is involved, this age may be an upper limit only. 

On closer inspection, we find that the change of argument of perihelion with node does not align all three bodies (Figure \ref{fig:orb_elements}). As noticed before by \citet{2017A&A...605A..68S}, the longitude of perihelion is not quite constant in the s-Taurids stream. Instead, the trend of argument of perihelion points directly at asteroid 2015 TX24, which is only 3.5 degrees further in node than the range of DFN fireballs observed. A few EN fireballs cover the node of 2015 TX24. 2025 TX24 is also close in position to the nodal dependence seen with perihelion distance, eccentricity and inclination, but those parameters can scatter more easily and that difference may merely reflect the change in orbital elements needed to intersect Earth's orbit. 

Asteroid 2005 UR and 2005 TF50 have similar longitude of perihelion as 2015 TX24, but don't fall along the trend line seen among the meteoroids. We conclude that these larger asteroids are not simply fragments of this stream as proposed by \citet{2017A&A...605A..68S}. 

A likely scenario is that we are seeing recent activity from a breakup involving surviving asteroid 2015 TX24. 2015 TX24 now has a semi-major axis of 2.2647 AU (Epoch 2020-Dec-17.0 TDB). That puts it within the normal semi-major axis oscillation of $\pm0.03$ AU from the resonance. Solar radiation pressure will slightly increase the semi-major axis of the DFN meteoroids, but if the ejection conditions were slightly lowering the semi-major axis, this could have been compensated. It is possible that the change of longitude of perihelion away from the node of 2015 TX24 is due the influence of the 7:2 mean-motion resonance. Presumably, the further the meteoroid node now is from 2015 TX24, the stronger the influence of the resonance. The precision of the semi-major axis is not good enough to verify that. If so, that could mean that the breakup happened rather recently, perhaps as recently as a few centuries ago.

This supports a model for the Taurid Complex showers that involves an ongoing fragmentation cascade of comet 2P/Encke siblings that were created following a breakup some 20,000 years ago (Jenniskens, 2006). In this scenario, 2015 TX25 broke with 2005 UR and 2005 TF50 from a larger precursor body about 1,500 years ago, and in the past few centuries a further breakup of 2015 TX24 created the fragments observed as the s-Taurids today. Asteroid 2003 WP21 does not belong to this group and was created earlier.

It is more difficult to understand how we could be seeing activity from comet 2P/Encke. In that case, the resonance must have more dramatically changed the nodal line. Comet 2P/Encke would be expected to cause meteoroid activity centered on $\lambda_{\odot} = 224.6\degr$, using method "H" of \citet{1990PASJ...42..175H}, implemented by \citet{1998A&A...331..411N}. However, Encke now has nodes close to perihelion near Mercury and aphelion in the asteroid belt. Comet Encke's orbit has changed its longitude of perihelion significantly over the past two centuries. In 1769, the node was at 157.46\degr, Pi = 159.19\degr. In 2020, the node is at 154.55\degr, Pi = 161.11\degr.
That rate of change is about the same as seen in the meteor stream. Around A.D. 1700, the longitude of perihelion of Encke was the same as that of the core of the s-Taurids. If we are seeing meteoroids ejected from 2P/Encke in the century around 1700 A.D., there must have been a subsequent dramatic change of the nodal line.

It is not clear whether or not the periodic nature of the s-Taurids is due to the 7:2 mean-motion resonance with Jupiter, other than preventing close encounters with Jupiter. The mean semi-major axis of the meteoroids appears to be related to that of its parent body. It is perhaps possible that the dust was released with very low relative ejection speeds and has remained concentrated in range of mean anomaly due to its young age. In particular, it is interesting that \citet{2017A&A...605A..68S} noted that the 1995 detected Taurid fireballs did not quite have the orbital elements of the 2015 Taurids. They had larger semi-major axes and smaller perihelia that did not change so much with solar longitude. This could point to the presence of streams being more dispersed but narrow, which occasionally wander in Earth's path. 

\subsection{Meteorite dropping Taurids?}
\citet{2013M&PS...48..270B} identified the Taurid showers as a potential source of macroscopic meteorite dropping events if a large enough meteoroid enters Earth's atmosphere. We have seen that the STS branch contains large members, do members of that population have a chance of surviving entry and falling as a meteorite? Large s-Taurids behave like weak matter (Fig. \ref{fig:heights}). They experience catastrophic disruption at very high altitudes ($>66$\,km, see suppl. mat. table). Their weak nature is not compensated by size, as \citet{2017A&A...605A..68S} noted the larger s-Taurids tend to be the weaker ones. The deepest penetrating STS observed by the DFN ($DN151114\_04$) is not visible below 52\,km. According to the criteria of \citet{2013M&PS...48..270B}, which states that a height of 35\,km and velocity of 10\,$\mbox{km s}^{-1}$ are approximate terminal dynamical criteria for a given event to have a chance of producing a meteorite fall, this is unlikely to produce a recoverable meteorite on the ground. The two very bright STSs described by \citet{2016MNRAS.461..674O,2017A&A...605A..68S} also terminate at high altitudes of 57.86 and 60.20\,km. 

Is this weakness a feature of all Southern Taurids? The deepest penetrating Southern Taurid (MORP $\#715$) described in the MORP dataset \citep{1996M&PS...31..185H}, only penetrates to 54.8\,km. As outlined by \citet{2013M&PS...48..270B}, one of the EN fireballs in 1995 penetrated as deep as 30\,km. Although this fireball was tentatively linked with the Taurid Complex, no definite association with either branch of the Taurids was reported, and the final velocity was not reported either. More generally, to our knowledge there is no report in the literature of a Southern Taurid that comes close to the terminal parameter of \citet{2013M&PS...48..270B}. 

On the other hand, we have examples of Northern Taurids that are able to penetrate much lower than the 50\,km ceiling that Southern Taurids seem to hit. For example, on October 9th, 2016, the DFN observed a Northern Taurid penetrating as deep as 36.4\,km, slowing down to 9.7\,$\mbox{km s}^{-1}$: the terminal parameters for this NTA are much closer to the cut-off criteria of \citet{2013M&PS...48..270B}, but still greater than required for a meteorite to survive. Are we seeing inhomogeneities here in the original composition of 2P/Encke? Or are the stronger materials representative for older meteoroids that survived the harsh conditions in the interplanetary medium?

\section{Conclusions}

The periodic outbursts of Taurid fireballs and visible meteors are from a stream called the s-Taurids (IAU shower 628, STS). The shower stands out well as a concentration of orbits in speed versus solar longitude diagram, with the shower members having a strongly changing entry speed with position along Earth's orbit. 

We have established the size frequency distribution for the s-Taurid stream. Even at gram sizes, the stream shows a very shallow distribution with a magnitude distribution index of 2.0 (differential mass distribution index of s = 1.75), atypical of other JFC showers. The distribution appears to remain unchanged up to metre-sized fragments.

The highly stratified structure of this stream, and the shallow size-frequency distribution remaining constant over a large range in size, points to the stream being the product of a gentle and relatively recent break up. 

Because the meteoroids initially move on different orbits than 2P/Encke, the breakup involved a different parent body. That body consisted of weak material. The distribution of longitude of perihelion along Earth's orbit points to the stream having originated from surviving asteroid \textit{2015 TX24} a low albedo 0.07, $0.25\pm0.04$ km sized ($H=21.5$) asteroid that is a good candidate for a 2P/Encke sibling, together with \textit{2014 TG10} and other such bodies identified earlier. 2015 TX24, 2005 UR and 2005 TF50 may have broken from a common body 1,500 years ago, with recent activity from 2015 TX24 now producing the meteoroids detected at Earth as the s-Taurids.

Large metre-sized bodies survive in the s-Taurids, possibly because they represent relatively strong cometary materials. During the s-Taurid \#628 stream (STS) outburst years, the chance of the Earth being hit by a metre scale weak meteoroid is enhanced by a factor of at least $[2.5,55]$. The Earth encounters the STS stream on average every 5 years, therefore the STS stream is responsible for as much as 20\% of all weak (airburst $>60\,$km altitude) metre scale bodies.

From the analysis of terminal parameters (heights and speeds) of the large Taurid meteoroids observed by the DFN, a macroscopic meteorite from a Southern Taurid stream seems unlikely, on the other hand we have examples of Northern Taurids that approach the meteorite dropping terminal parameters discussed by \citep{2013M&PS...48..270B}. Given their shallow penetration depth, large meteoroids from the s-Taurids are unlikely to drop meteorites, but might generate dust that can be collected in the atmosphere.

When the cloud of meteoroids next returns close to the Earth in 2022, the Large Synoptic Survey Telescope \citep{2008SerAJ.176....1I} will be able up and running and should be able to better constraint on the hypothetical population of metre-sized s-Taurids.  The recently commissioned Geostationary Lightning Mapper \citep{2018M&PS...53.2445J} should also be able to fill the observation gap around the decimetre scale, and notably provide good estimate of the flux density of STS meteoroids in this range.

\begin{acknowledgments}
This research is supported by the Australian Research Council through the Australian Laureate Fellowships scheme, receives institutional support from Curtin University, and uses the computing facilities of the Pawsey supercomputing centre. The DFN data reduction pipeline makes intensive use of Astropy, a community-developed core Python package for Astronomy \citep{2013A&A...558A..33A}. PJ is supported by grant 80NSSC19K0513 of NASA's Emerging Worlds program and grant 80NSSC19K0563 of NASA's Solar System Workings program.
\end{acknowledgments}

\bibliography{research}{}
\bibliographystyle{aasjournal}

\end{document}